\documentclass[submission,copyright,creativecommons]{eptcs}
\providecommand{\event}{ICLP 2020} 
\usepackage{underscore}           

\newcommand{\as}{``}
\newcommand{\ass}{`}

\title{Logic Programming and Machine Ethics}
\author{Abeer Dyoub \quad Stefania Costantini
	\institute{{DISIM, }
	{University of L'Aquila, Italy}}
	\email{Abeer.Dyoub@graduate.univaq.it \quad Stefania.Costantini@univaq.it}
	\and
	Francesca A. Lisi
	\institute{DIB and CILA, University of Bari ``Aldo Moro", Italy}
	\email{\quad FrancescaAlessandra.Lisi@uniba.it}
}
\def\titlerunning{Logic Programming and Machine Ethics}
\def\authorrunning{A.\,Dyoub and S.\,Costantini and F.\,A. Lisi}

\begin{document}
\maketitle

\begin{abstract}
Transparency is a key requirement for ethical machines. Verified ethical behavior is not enough to establish justified trust in autonomous intelligent agents: it needs to be supported by the ability to explain decisions. Logic Programming (LP) has a great potential for developing such perspective ethical systems, as in fact logic rules are easily comprehensible by humans. Furthermore, LP is able to model causality, which is crucial for ethical decision making.
\end{abstract}

\section{Introduction}
\label{intro}

Autonomous Intelligent Systems are designed to reduce the need for human intervention in our daily life. However, the full benefit of these new systems will be attained only if they are aligned with society's values and ethical principles.
Adopting ethical approaches to building such systems has been attracting a lot of attention in the recent years. The global concern about the ethical behavior of this kind of technologies has manifested in many initiatives at different levels. As examples, we mention: the IEEE initiative for ethically aligned design of autonomous intelligent systems ('Ethics in Action'  \footnote{\url{https://ethicsinaction.ieee.org}}), and the European Commission famous 'Ethics Guidelines for Trustworthy AI'\footnote{\url{https://ec.europa.eu/digital-single-market/en/news/ethics-guidelines-trustworthy-ai}}. In the latter document, the European Commission's High-Level Expert Group on Artificial Intelligence (AI) specifies the requirements of trustworthy AI, and the technical and non-technical methods to ensure the implementation of these requirements into AI systems.

The tech giant 'Google', after a protest aroused by company employees over ethical concerns, ended its involvement in an American Pentagon Project on autonomous weapons\footnote{\url{https://www.nytimes.com/2019/03/01/business/ethics-artificial-intelligence.html}}. Because of the controversy over its Pentagon work, Google laid down its own set of AI ethical principles \footnote{\url{https://www.blog.google/technology/ai/ai-principles/}} meant as a guide for future projects. However, the new principles are open to context-related interpretations.

Overall, in spite of the significant amount of interest that \ass Machine Ethics' has received over the last decade mainly from ethicists and artificial intelligence experts, the question \as are artificial moral agents possible?'' is still lingering. Machine ethics has evolved as a new field aiming at creating machines able to compute and choose the best ethical action. The term \ass Artificial Moral Agent' (AMA) has been around since a long time, at least since the science fiction author Isaac Asimov stated the \as Three Laws of Robotics'' in a short story in 1942 and then in the famous book \as I, Robot'' \cite{asimovrobot}. For decades, many researchers have voiced their concerns about the need of machine ethics. 

Moral decision making and judgment is a complicated process involving many aspects: it is considered as a mixture of reasoning and emotions. In addition, moral decision making is highly flexible, contextual and culturally diverse. 
Since the beginning of this century there have been several attempts for implementing ethical decision making into intelligent autonomous agents, using different approaches. So far however, no fully descriptive and widely acceptable model of moral judgment and decision making exists. None of the developed solutions seem to be fully convincing to provide a trusted moral behavior. In addition, all the existing research in machine ethics tries to satisfy certain aspects of ethical decision making, but, maybe inevitably, fails to satisfy others.   

Approaches to machine ethics can be classified into two categories. Top-down approaches are those which try to implement some specific normative theory of ethics into autonomous agents so as to ensure that an agent acts in accordance with the principles of this theory. Bottom-up approaches are developmental or learning approaches, in which ethical mental models emerge via the activity of individuals rather than expressed explicitly in terms of normative theories of ethics \cite{WallachAS08}. In other words, generalism versus particolarism, principles versus case based reasoning. Both approaches have advantages and disadvantages. We therefore need hybrid approaches, able to combine techniques related to the two perspectives in one framework.

An important aspect of implementing machine ethics in intelligent agents involves trust and safety, i.e., agents should do what they are expected to do and should not cause harm. They should also be able to explain and justify their choice of a particular action over others (explainability/transparency), and be accountable for their actions (ethical responsibility). We can address these issues via certification and run-time assurance \cite{SSS1817486}, \ass on top' of approaches to implement ethics in agents that support explainability and accountability.

Machine Learning (ML) is currently used for critical prediction applications also in fields like healthcare and criminal justice, where their predictions have a deep impact on human lives (which clearly involves ethical issues). Most of these ML models are \as black boxes'', meaning that the model internal logic
and inner workings are hidden to the user. This is a serious disadvantage because it prevents a human,
expert or non-expert, from being able to verify, interpret, and understand the reasoning of the system
and how particular decisions are made \cite{MontavonLBSM17}. 
The lack of transparency and accountability of these predictive models can have severe consequences \cite{AdadiB18}. Examples of such systems are (among many): \cite{wexler2018life}, \cite{VarshneyA16}, \cite{mcgough2018bad}.

Ethical behavior, even if it were verifiable, is not enough to establish justified trust in an autonomous system. It needs to be supported with the ability to explain decisions. In other words, there is a need for a \as machine explainability' \footnote{In this work, we use the words explainability, interpretability, comprehensibility interchangeably.} component along with an ethical component. The explainability component is meant to enable autonomous systems to explain their actions and justify their decisions, so that human users can understand their behavior, and as a result trust them. 

One of the most important ultimate goals of explainable AI systems is the efficient mapping between explainability and causality, which is the key to effective human-AI interaction \cite{abs-1906-00657}. Explainability is the system's ability to explain itself in natural language to an average user by being able, e.g., to say: \as I generated this output because x,y,z''. In other words, the ability is required of a system to state the causes behind its decisions. Causality is central to ethical decision making \cite{beebee2009oxford}.

Logic Programming, meaning \ass classical' Prolog \cite{lloyd} and other related paradigms like Answer Set Programming (ASP, cf., among many, \cite{ASPJournal2016} and the references therein) has been used by many researchers to implement machine ethics. Logic-based approaches have a great potential to model moral machines, in particular via non-monotonic logics. This potential is due to many factors like: ethical theories and dilemmas have always been represented in a declarative form by ethicists, who also used formal and informal logic to reason about them. Furthermore, it is common understanding that ethical rules are default rules, which means that they tolerate exceptions. This fact nominates non-monotonic logics, which simulate commonsense reasoning, as suitable tools to formalize different ethical conceptions. 
ILP (Inductive Logic Programming) algorithms (cf.\cite{muggleton1991inductive,LawRB19,GulwaniHKMSZ15}) are a subclass of ML algorithms aimed at learning logic programs. ILP does not require huge amounts of training examples such as other (statistical) Machine Learning methods and produces interpretable results, that means a set of rules which can be analyzed and adjusted if necessary. So, ILP appears to be a suitable and promising technique for implementing machine ethics, where scarcity of examples is one of the main challenges, and comprehensibility of the output is indispensable.

Classical statistical ML approaches (also called \ass feature-based learning') are in fact limited to non-relational descriptions of objects: i.e., the learned descriptions (predictions) do not specify the relations among the objects parts. According to \cite{BratkoM95}, classical ML approaches have two strong limitations, namely, i) the background knowledge can be expressed in rather limited form; ii) the lack of relations makes the concept description language inappropriate for some domains.
Standard supervised machine learning algorithms for regression and classification are inherently based on finding correlations in the data. These correlations stand for a probability that things will turn out the same in the future. What they do not reveal is why this should be the case \cite{Hildebrandt08}. In other words, there is no care for causal reasoning or \ass explanation' beyond the statistical sense. Instead, ILP produces comprehensible logical rules (inference rules) in which the head of the rule (the conclusion) is inferred from the body's parts of the rule, in other words, the body (premises) causes the head (conclusion). Comprehensibility of logic-based representations is in fact one of their most recognized advantages.

In this work, after providing some background and reviewing the state of the art of logic-based approaches concerning machine ethics, we present our proposal for constructing ethical machines, which is a hybrid logic-based approach. All along, we will try to emphasize the great potential of logic programming for implementing ethical machines. 

\subsection{ASP and ILP in a Nutshell}
\label{pre}
\subsubsection{Answer Set Programming (ASP) and Prolog}
ASP \cite{ASPJournal2016} is a logic programming paradigm under the answer set (or  \as stable model'') semantics \cite{GelLif88}, which applies ideas of autoepistemic logic and default logic. In ASP, search problems are reduced to computing the answer sets of a given ASP program via an \ass answer set solver', i.e., an inference engine for generating stable models.
An ASP program is a collection of rules of the form: $H\leftarrow A_{1} , \ldots , A_m, not\, A_{m+1}, \ldots, not\, A_n$ 
were each of the $A_i$'s is a literal in the sense of classical logic. Intuitively, the above rule means that if $A_1, \ldots , A_m$ are true and if $A_{m+1}, \ldots , A_n$ can be safely assumed to be false, then $H$ will be true. The left-hand side and right-hand side of rules are called \emph{head} and \emph{body}, respectively. A rule with empty body ($n = 0$ and $m = 0$) is called a \emph{fact}. A rule with empty head is a \emph{constraint}, and states that literals of the body cannot be simultaneously true in any answer set. A program may have several answer sets or may have no answer sets at all. Thus, a program $\Pi$ describes a problem, of which its answer sets represent the possible solutions. In case the program involves no constraints and no cycles on negation, an ASP program coincides with a Prolog program \cite{lloyd}, that can be run and queried by means of one of the many available Prolog interpreters\footnote{Many performant ASP solvers an Prolog interpreters are freely available, a list of them is reported at \url{https://en.wikipedia.org/wiki/Answer_set_programming} and \url{https://en.wikipedia.org/wiki/Comparison_of_Prolog_implementations} respectively.}.
\subsubsection{Inductive Logic Programming (ILP)}
ILP \cite{muggleton1991inductive,LawRB19,GulwaniHKMSZ15} is a branch of Artificial Intelligence which investigates the inductive construction of logical theories from examples and background knowledge. In the general setting, we assume a set of Examples \textit{E}, positive $E^+$ and negative $E^-$, and some background knowledge \textit{B}. An ILP algorithm finds the hypothesis \textit{H} such that $B \bigcup H \models E^+$ and $B \bigcup H \not\models E^-$. The possible hypothesis space is often restricted via a language bias that is specified by a series of mode declarations \textit{M}. A mode declaration is either a head declaration \textit{modeh(r, s)} or a body declaration \textit{modeb(r, s)}, where \textit{s} is a ground literal, this scheme serves as a template for literals in the head or body of a hypothesis clause, where \textit{r} is an integer, the recall, which limits how often the scheme can be used. A scheme can contain special
\textit{placemarker} terms of the form \textit{$\sharp$ type}, \textit{+type} and \textit{-type}, which stand, respectively, for ground terms, input terms and output terms of a predicate \textit{type}. Finally, it is important to mention that ILP has found applications in many areas. For more information on ILP and applications, cf., among many, \cite{MuggletonR94} and references therein.

\section{Logic for Programming Ethical Machines: State of the Art}
\label{state}


Tom Powers in \cite{powers2006prospects} assesses the viability of using deontic and default logics, to implement Kant's categorical imperative.
Kant's categorical imperative (\ass Act only according to that maxim, whereby you can at the same time, will that it should become a universal law without contradiction' \cite{paton1971categorical}).
Three views on how to computationally model the categorical imperative have emerged in these works.
First, in order for a machine to maintain consistency in testing ethical behavior, one should construct a moral theory for individual maxims, and map them onto deontic categories. Deontic logic is regarded as an appropriate formalism with respect to this first view.
Second,	there is the need for commonsense reasoning in the categorical imperative, to deal with contradiction. Under this view, non-monotonic logic can be appropriate to capture defeating conditions to a maxim: Reiter's default logic \cite{reiter1980logic}
has been in fact regarded as a suitable formalism. Third, the construction of a coherent system of maxims, where the authors see such construction as analogous to the belief revision problems. In the context of bottom-up construction, an update procedure is envisaged for a machine to update its system of maxims with another maxim, though it is unclear how such an update can be accomplished. These three views were only considered abstractly, and no implementation has been proposed to address them.

In \cite{BringsjordAB06}, the authors suggest that mechanized multi-agent deontic logics might be an appropriate tool for engineering ethically correct robot behaviors. They use the logical framework Athena \cite{arkoudas2005toward}, to encode the natural deduction system of Murakami \cite{Murakami04} as an axiomatization of Horty's utilitarian formulation of multi-agent deontic logic \cite{horty2001agency}. The use of an interactive theorem prover is motivated by
the idea that agents operate according to ethical codes bestowed on them, and when their automated reasoning should fail, they should suspends their operation and ask humans' guidance to resolve the issue.
Making an example in health care, where two agents are in charge of two patients
with different needs (patient H1 depends on life support, whereas patient H2 on very costly pain medication), two actions are considered: (1) terminate H1's life support to secure his organs for five humans; and (2) delay delivery of medication to H2 to preserve hospital resources. The example then considers several
candidate ethical codes, from harsh utilitarian (that both terminates H1's life and delays H2 medication) to most benevolent (which neither terminates H1's life nor delay H2 medication); these ethical codes are formalized using the aforementioned deontic logics. The logic additionally formalizes behaviors of agents and their respective moral outcomes. Given these formalizations, Athena is employed to query each ethical code candidate in order to decide which amongst them should be operative.

Other attempts tried to formalize ethical systems using modal logic formalisms \cite{gensler1996formal} and then trying to operationalize these formalizations, like in \cite{BringsjordAB06} and \cite{powers2006prospects}. These formalizations are mainly based on the use of deontic logics \cite{meyer1994paradoxes}, that are suitable for ethical systems focused on laws where permission and prohibitions are well defined, but not to consequentialist ethical systems, that judge whether or not something is right by what its consequences are.

Pereira (with several co-authors) has proposed the use of different logic-based features for representing in logic programming diverse issues of moral facets, such as moral permissibility, doctrines of Double Effect and Triple Effect, the Dual-process Model, counterfactual thinking in moral reasoning. They investigate the use of abduction, probabilistic logic programming, logic programming updating, tabling. These logic-based reasoning features have been synthesized in three different systems: ACORDA, Probabilistic EPA, QUALM (cf., among many, \cite{PereiraS16,SaptawijayaP16,PereiraL20} and the references therein).

One way of implementing ethical decision making is qualifying and quantifying the good and the bad ramifications of ethical decisions before taking them. This task is non-trivial, as there could be many approaches for doing this. First, qualifying the \ass Good' involves identifying modes for defining this concept, which is a controversial task because there exist a lot of theories attempting to define notions of \ass Good'. \cite{BerrebyBG17} presents a model for quantifying the good after it has been qualified. For qualifying the good, they present two modes, one based on rights and the other one based on values. For quantifying the good, they propose a method in which they define three weighing parameters for the good and the bad ramifications of events caused by actions. Then, they integrate all weights into a single number, which represents the weight of an event in relation to a particular modality and group of people. The total weight of an event then is the difference between the sums of all its weighted good and bad ramifications. Greater weights correspond to more participation in the good, while negative weights do more harm than good. Their approach was implemented in ASP.
 
In \cite{ganascia2007modelling}, the authors formalized three ethical conceptions (the Aristotelian rules, Kantian categorical imperative, and Constant's objection) using nonmonotonic logic, particularly Answer Set Programming. 



 In \cite{CointeBB16}, authors introduced a model that can be used by an agent in order to judge the ethical dimensions of its own behavior and the behavior of others. Their model was implemented in ASP. However, the model is still based on a qualitative approach. Whereas it can define several moral valuations, there is neither a degree of desires, nor a degree of capability, nor a degree of rightfulness. Moreover, ethical principles need to be more precisely defined to capture various sets of theories suggested by philosophers.
 
Sergot in \cite{sergot2016engineering}, provides an alternative representation to the argumentative representation of a moral dilemma case concerning a group of diabetic persons, presented in \cite{AtkinsonB06}, where the authors used value-based argumentation to solve this dilemma. According to Sergot, the argumentation framework representation does not work well and does not scale. Sergot proposal for handling this kind of dilemmas is based on Defeasible Conditional Imperatives. The proposed solution has been implemented in ASP.

JEREMY \cite{anderson2004towards} is an implementation of the Hedonistic Act Utilitarianism. This theory states that an action is morally right if and only if that action maximizes the pleasure, i.e. the one with the greatest net pleasure consequences, taking into account those affected by the action. The theory of Act Utilitarianism has, however, been questioned as not entirely agreeing with intuition.
The authors of JEREMY, to respond to critics of act utilitarianism, have created another system, W.D. \cite{anderson2004towards} which avoids a single absolute duty, by following several duties. Their system follows the theory of \ass prima facie' duties of Ross \cite{ross2002right} and is implemented via ILP.
Ethics is more complicated than following a single ethical principle. According to Ross (\cite{ross2002right}), ethical decision making involves considering several prima facie duties, and any single-principled ethical theory like Act Utilitarianism is sentenced to fail.

ILP has been used by researchers to model ethical decision making in MedEthEx \cite{AndersonAA05}, and EthEl \cite{AndersonA08}. These two systems are based on a more specific theory of prima facie duties viz., the principle of Biomedical Ethics of Beauchamp and Childress \cite{beauchamp1991principles}.
In these systems, the strength of each duty is measured by assigning it a weight, capturing
the view that a duty may take precedence over another. Then, for each
possible action, the weighted sum of duty satisfaction is computed, and the \ass right' action is the one with the greatest sum. The three systems use ILP to learn the relation \textit{supersedes(A1,A2)} which says that action \textit{A1} is preferred over action \textit{A2} in an ethical dilemma involving these choices. 
MedEthEx is designed to give advice for dilemmas in biomedical fields, while EthEl is applied to the domain of eldercare with the main purpose to remind a patient to take her medication, taking ethical duties into consideration. 
GenEth \cite{AndersonA14} is another framework that makes use of ILP. GenEth has been used to codify principles in a number of domains relevant to the behavior of autonomous systems.

\section{Why Logic Programming for Machine Ethics}
\label{why}

As seen in previous section, Logic Programming, and particularly Answer Set Programming and Inductive Logic Programming, have been used to formalize different ethical conceptions, where logical representations help to make ideas clear and highlight features, advantages and disadvantages of different ethical systems.
In contrast, standard ML (black-box) classifiers have many drawbacks, including \cite{abs-1911-01547,abs-1801-00631}:
poor generalization;
being intransparent (opaque), which makes the evaluation of safety of a software involving such classifiers problematic. This causes experts as well as non-expert users hesitate to trust such systems. Moreover, the
training process needs a huge number of training examples. 

Comprehensible (interpretable) models may help improve the user trust in the model, particularly when the system (algorithm) produces unexpected results (output/model) \cite{Lavrac99}.
A real-world example related to the problem of trust in ML systems is the terrible accident that happened in the Tree-Mile island nuclear power plant. There, the human operator did not implement the shutdown recommended by the plant's automated system, because she/he did not trust the system's recommendation \cite{henery1995}.

ILP has shown a great potential for addressing the aforementioned limitations of standard ML approaches, and complements deductive programming approaches \cite{BodikT12,MannaW80}.
ILP provides many advantages compared to standard (statistical) ML approaches:

\begin{itemize}
	\item ILP systems can learn complex relational theories due to the expressiveness of logic programs.
	\item ILP systems can learn by exploiting Background-Knowledge (BK).
	\item ILP systems can generalize from a small number of examples, thanks to the use of BK as a form of inductive bias. BK is similar to features used in most forms of ML \cite{CropperDM20}, and it is even more general as it may include rules, constraints and general principles.
	\item Learned hypotheses are logic programs, so they are comprehensible by humans.	
	\item ILP systems naturally support lifelong and transfer learning \cite{Cropper20}. Because of the symbolic representation of the learned knowledge which can be remembered by explicitly storing it in the BK. This fact is considered fundamental for human-like AI \cite{LakeUTG16}.
	\item In cases where it is hard for human inductive reasoning to synthesize a specific algorithm's details, ILP can be used to induce program candidates from user-provided data or test cases \cite{SousaSDPGGSH16}.
\end{itemize} 


In many application domains, especially when human lives are involved, users need to understand well and to have trust in the system's recommendations, so as to be able to explain the reasons for their decisions to other people. For example, in the medical domain, if a doctor recommends a surgery for his patient based on the prediction of an ML model, and this decision may cause major harm to the patient, the doctor needs to understand properly the reasons behind the ML model's predictions, in order to be able to defend herself in the court if sued for medical negligence \cite{richards2001data}. Another example where legal explanations are needed is the credit scoring applications, where banks are often legally obliged to explain to a customer why she/he was denied a credit \cite{MartensVVB11}.

Reviewing the literature on explainability (interpretability), there exist two main paths to achieve interpretability:
creating inherently-interpretable models by design (which is our suggested approach);
creating explanation methods for explaining black box, opaque models (post-hoc explanations).

It is worth noting that most methods for generating post-hoc explanations are themselves based on statistical tools that are subject to uncertainty or errors.
Many of the post-hoc interpretability techniques try to approximate deep-learning black-box models with simpler interpretable models that can
be inspected to explain the black-box models. However, these approximate models are often not loyal with respect to the original model, as there are always trade-offs between explainability and fidelity.
Typical approaches of the first path are decision trees and variants such as decision sets \cite{LakkarajuBL16}. 
However, decision trees generalize over feature vectors, and concepts are expressed as disjunction of conjunctions of constraints over features. ILP offers a more expressive approach: in ILP in fact, training examples are characterized by relations and the learned hypotheses are represented as logical rules, thus, relational and even recursive (infinite) concepts can be learned. Learned hypotheses are represented on the symbolic level, and so they are inspectable and comprehensible \cite{GulwaniHKMSZ15}.

In \cite{MichieDonald}, Donald Michie defined criteria for Machine Learning performance. 
His criteria incorporate two dimensions, viz. Predictive Accuracy and Comprehensibility of the generated knowledge. Unfortunately, later definitions tended to concentrate on one dimension, viz. predictive accuracy, which is readily measurable, neglecting comprehensibility which is not easily quantifiable, thus ultimately favoring statistical over symbolic Machine Learning approaches. Michie provided a three level criteria: weak, strong, and ultra strong.
The weak criterion is when the predictive performance of the learner improve with more amounts of data. The strong one has the further requirement to provide the generated knowledge in a symbolic form. Finally, the ultra strong criterion extends the strong one by adding that the learner system should teach the learned hypothesis to a human, which would lead to a better performance compared to the human studying the data on her/his own.

Most of current ML systems meet the weak criterion, while ILP meets the strong one. Furthermore, lately the authors in \cite{MuggletonSZTB18}, while attempting to meet the ultra strong criterion, provided an operational definition of comprehensibility of a generated hypothesis which can be estimated using human participant trail. Then, they conducted experiments in which they showed that participants were not able to learn the relational concept on their own from the dataset, but they were able to apply the relational definition generated by the ILP system correctly. This result demonstrates that ILP systems can fulfill Michie's criterion of operational effectiveness.
As discussed, e.g., in \cite{GulwaniHKMSZ15}, there are domains and problems in criminal justice, healthcare, and computer vision, where interpretable models such as ILP could profitably replace black box models.

\section{Our Proposal for building Ethical Agents}
\label{proposal}
Autonomous machines (agents) are increasingly engaging in human communities. Thus, agents should be expected to follow the community's social and ethical norms. Embedding norms in such autonomous systems requires a clear outlining of the community in which they are to be deployed. In fact, determining which moral values to aim for and which ethical principles to adhere to in given circumstances is one of the main challenges for ethical reasoning. Codes of ethics and conduct provide us with a framework to work within. 
However, these codes are based upon general principles such as confidentiality, accountability, honesty, inclusiveness, empathy, fidelity, etc., that are quite difficult to put into practice in their abstract form \cite{jonsen1988abuse}. Moreover, abstract principles such as these may contain terms whose meaning may change according to the context. It is difficult to use deductive logic only to address such a problem: it is in fact hardly possible for experts to define fine-grained detailed rules to cover all possible situations. 

We need to teach our machines the codes of ethics and conduct of the domain in which they are to be deployed. We believe that Artificial Agents could, similarly to humans, acquire ethical decision making and judgment capabilities by iterative learning processes, in particular inductive learning \cite{WallachAS08}.
With increasing autonomy, there will be in fact more situations that require morally relevant decisions to be made. Many of these decisions cannot be foreseen in advance in their full detail.
Therefore, we need bottom-up (learning) approaches because it is difficult to fully specify in advance all possible scenarios (framing problem), and because there is no actual agreement about which explicit theory of normative ethics should be implemented \cite{ConitzerSBDK17}.

Our approach to implementing ethical agents combines deductive (rule-based) logic programming and inductive (learning) logic programming approaches in one framework. We use ASP for knowledge representation and reasoning, and ILP as a machine learning technique for learning from cases and generating the missing detailed ethical rules needed for reasoning about future similar cases. The newly learned rules are to be added to the agent's knowledge base. ASP was chosen as a suitable tool for formalizing different ethical conceptions, because ethical rules are known to be default rules, which means that they tolerate uncertainty and exceptions. In addition, there are the many advantages of ASP including it is expressiveness, flexibility, extensibility, ease of maintenance, readability of the code. The availability of performant solvers can help to perform comparisons of consequences of ethical theories, and makes it easier to validate our models in different situations. ILP was chosen because, as a logic-based machine learning approach, it supports two very important and desired aspects of machine ethics implementation into artificial agents, viz. explainability and accountability, and it is known for its explanatory power. Moreover, ILP can be better suited than statistical methods whenever training examples are scarce, as it is often the case in domains involving ethical aspects.

The application domain that we considered as a case study is that of online customer service chatbots. In fact, 
codes of ethics in this domain are abstract general principles, that apply to a wide range of situations. They are subject to interpretations and may have different meanings in different contexts. There are no intermediate rules that elaborate these abstract principles or explain how they apply to concrete situations. 
Our system is able to learn new ethical evaluation rules according to: facts and ethical evaluation provided by a trainer, and a background knowledge base. In particular, we exploited ILED, one of the state-of-the-art ILP tools for learning ASP programs. For lack of space we cannot illustrate here the technical aspects of our approach, so we refer the interested readers to \cite{ADILP2019,ADSCFL2019}. 

\section{Discussion and Conclusions}
\label{conclude}

In the context of many application domains and \emph{a fortiori} in domains involving ethical aspects, it is crucial that systems' decisions are transparent and comprehensible and in consequence trustworthy.
Comprehensibility is one of the main features that distinguish logic-based representations from those proper of statistical ML.
This makes ILP appropriate for scientific theory formation tasks in which the comprehensibility of the generated knowledge is essential. Moreover, ILP is able to learn rules from a small number of examples. This makes it suitable for ethical domains, where scarcity of examples is one of the main challenges. 

Combining logic-based representation and logic-based learning for modeling ethical agents, as done in our aforementioned work, provides many advantages: increases the reasoning capability of agents; promotes the adoption of hybrid strategies that allow both top-down design and bottom-up learning via context-sensitive adaptation of models of ethical behavior; allows the generation of rules with valuable expressive and explanatory power, thus equipping agents with the capacity to make ethical decisions, and to explain the reasons behind these decisions. 
In our opinion and for the sake of transparency, ethical decision-making and judgment should however be guided by explicit ethical rules determined by competent judges or ethicists, or generated automatically but approved through consensus of ethicists. 

In conclusion, we believe that logic-based approaches, which are inherently-interpretable, have a great potential for implementing ethical machines, avoiding the potential problems caused by black-box ML models. This especially in consideration of the recent advances in Inductive Logic Programming \cite{CropperDM20} which puts it in a position to substitute black-box machine learning, particularly in critical applications. An interesting immediate future direction for our work is exploitation of the results of \cite{abs-2005-00904} which proposes a new tool, ILASP, for learning ASP program fragments.

\end{document}